\documentclass[[%
iopart,
amsmath,amssymb,
]{revtex4-1}

\usepackage{placeins}
\usepackage{subcaption}
\usepackage{graphicx}
\usepackage{dcolumn}
\usepackage{bm}
\usepackage{xcolor}

\begin{document}


\title{Laser-Cooled Polyatomic Molecules for Improved Electron Electric Dipole Moment Searches}

\author{Benjamin L. Augenbraun$^{1,2}$}
\email{augenbraun@g.harvard.edu}
\author{Zack D. Lasner$^{1,2}$}
\author{Alexander Frenett$^{1,2}$}
\author{Hiromitsu Sawaoka$^{1,2}$}
\author{Calder Miller$^{1,2}$}
\author{Timothy C. Steimle$^{3}$}
\author{John M. Doyle$^{1,2}$}
\affiliation{$^1$Department of Physics, Harvard University, Cambridge, MA 02138, USA}
\affiliation{$^2$Harvard-MIT Center for Ultracold Atoms, Cambridge, MA 02138, USA}
\affiliation{$^3$School of Molecular Science, Arizona State University, Tempe, AZ 85287, USA}

\date{October 23, 2019}

\begin{abstract}
Doppler and Sisyphus cooling of $^{174}$YbOH are achieved and studied. This polyatomic molecule has high sensitivity to physics beyond the Standard Model and represents a new class of species for future high-precision probes of new T-violating physics. The transverse temperature of the YbOH beam is reduced by nearly two orders of magnitude to $< 600 \, \mu$K and the phase-space density is increased by a factor of $>6$ via Sisyphus cooling. We develop a full numerical model of the laser cooling of YbOH and find excellent agreement with the data. We project that laser cooling and magneto-optical trapping of long-lived samples of YbOH molecules are within reach and these will allow a high sensitivity probe of the electric dipole moment (EDM) of the electron. The approach demonstrated here is easily generalized to other isotopologues of YbOH that have enhanced sensitivity to other symmetry-violating electromagnetic moments.
\end{abstract}

\maketitle

\section{Introduction}
Experimental probes of the electric dipole moment of the electron (eEDM) provide strong constraints on theories of particle physics beyond the Standard Model (BSM)~\cite{Bernreuther1991, Pospelov2005, Engel2013, DeMille2015, Nakai2017, Cesarotti2019}. The most stringent limit on an eEDM has been realized in experiments using a diatomic molecule, ThO, in a $^3 \Delta_1$ state, limiting the eEDM to $<1.1 \times 10^{-29} \, e\text{ cm}$~\cite{ACME2018, ACME2014}. This has placed limits on T-violating new physics above the TeV scale~\cite{Nakai2017,Cesarotti2019}. Other work, using HfF$^+$ with the same electronic structure, has confirmed the ACME results at the $<1.3 \times 10^{-28} \, e\text{ cm}$ level~\cite{Cairncross2017}. The sensitivity of these experiments comes in part from the particular structure of the angular momentum states in these molecules. Specifically, orbital angular momentum along the internuclear axis allows one to fully polarize the molecules in the lab frame, thereby providing control of the large internal effective electric field ($>$10-100~GV/cm)~\cite{Petrov2007,Skripnikov2013}. For these two molecules, this structure and concomitant ease of polarization, which is the result of closely spaced levels of opposite parity in $\Omega$-doublet states, allows for strong rejection of many systematic errors~\cite{Meyer2008}. A future eEDM search could combine this internal structural feature with other advances, such as extended coherence times and larger numbers of molecules. It was recently proposed~\cite{kozyryev2017PolyEDM} that polyatomic molecules generically allow for eEDM searches that combine scalability, polarizability, long coherence times, and robustness to systematic errors. In particular, YbOH was pointed out as a viable candidate for greatly improved searches of symmetry violating new physics. Polyatomic molecules are also of interest to other tests of BSM physics. For example, degenerate bending modes can provide avoided crossings useful in probing nuclear spin-dependent parity violation~\cite{Norrgard2019}.

As described in Ref.~\cite{kozyryev2017PolyEDM}, eEDM-sensitive molecules with nonzero angular momentum projection along the internuclear axis arising from electronic orbits, i.e. the highly polarizable $^3\Delta_1$ state, have very poor laser cooling properties. No diatomic molecule with sensitivity to new physics has been identified that is simultaneously amenable to laser cooling while also providing convenient parity doublets for polarization and systematic error control. Polyatomic molecules, on the other hand, generically possess small parity doublets arising from nuclear orbital motion. For example, the low-lying bending modes of linear triatomic molecules in $^2\Sigma$ states posses nearly degenerate levels of opposite parity that are metastable. A class of molecules has been identified where a precision measurement with trapped polyatomic molecules, e.g. $^{174}$YbOH, could probe CP-violating BSM physics at the PeV scale in a near-term experiment~\cite{Prasannaa2019,kozyryev2017PolyEDM}.

In order to achieve long coherence times and manage systematic errors, trapping of molecules under weakly perturbing conditions and with long lifetimes, e.g. in an optical dipole trap (ODT), will be crucial in the pursuit of next-generation eEDM measurements. Weak traps require much lower molecule temperatures than were achieved in recent eEDM experiments, and therefore some form of deeper cooling is necessary. Direct laser cooling of molecules~\cite{dirosa2004laser,stuhl2008magneto,Carr2009, McCarron2018,Tarbutt2018} has seen rapid growth in recent years; SrF~\cite{Shuman2009,Shuman2010, Barry2012, Barry2014, McCarron2015, norrgard2015sub, Steinecker2016, McCarron2018}, CaF~\cite{Truppe2017b, truppe2017CaF, anderegg2017CaFMOT,  Williams2017, Cheuk2018lambda, Caldwell2019}, and YO~\cite{Hummon2013,Yeo2015,Collopy2018} have all been laser-slowed, cooled, trapped, and transferred to long-lived traps. YbF~\cite{Lim2018} molecules have been transversely laser-cooled, although not yet cooled in three dimensions. In experimental work begun in 2014, the polyatomic radical SrOH was sub-Doppler cooled in one dimension~\cite{kozyryev2016Sisyphus}, marking a path to cooling of much heavier and more complex isoelectronic species, like YbOH. To date, there has been no experimental demonstration of laser cooling of a polyatomic molecule sensitive to the eEDM. This is in part because heavy polyatomics, like YbOH, are significantly more challenging than previously cooled species due to their high masses, strong perturbations in the electronically excited states, and less favorable Franck-Condon factors.

We report here one-dimensional Doppler and Sisyphus cooling of a beam of $^{174}$YbOH from 20~mK to below 600~$\mu$K. The particular Sisyphus effect used here, called the magnetically-assisted Sisyphus effect, has been investigated previously in both atoms~\cite{Sheehy1990,Emile1993,Kloter2008} and molecules~\cite{Shuman2010,kozyryev2016Sisyphus,Lim2018}. Consistent with previous work on atoms and lighter molecular species, we observe that Sisyphus cooling is more efficient than Doppler cooling because the magnitude of the cooling force is set by the depth of the laser field's light shifts, which can be made arbitrarily deep at high intensity~\cite{CohenTannoudji1992,Emile1993,Phillips1998}. The laser cooling demonstrated here is a crucial proof-of-principle test for further direct cooling and trapping of YbOH molecules for a new generation of eEDM experiments. We create a theoretical simulation based on the optical Bloch equations and find excellent agreement with our experimental data. We discuss extensions of our approach to isotopologues of YbOH, including $^{173}$YbOH, which have been proposed for use in measurements of the nuclear magnetic quadrupole moment~\cite{Maison2019}.

\section{Experiment}
\begin{figure}[t]
\begin{centering}
\includegraphics[width=0.8\columnwidth]{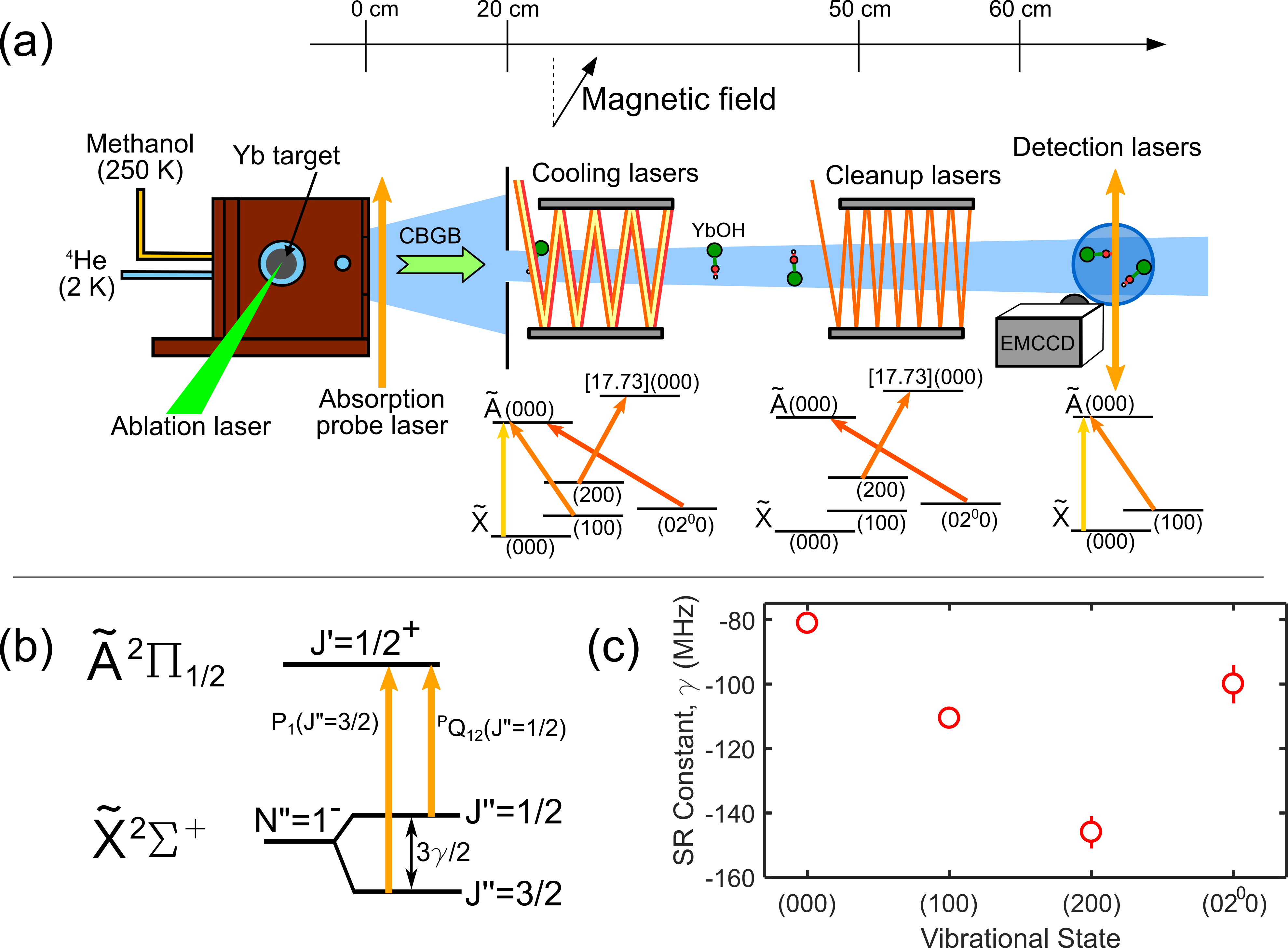}
\par\end{centering}

\caption{\label{fig:fig-Setup} (a) Schematic of the experimental
apparatus (not to scale). A cryogenic beam of YbOH is produced by ablating a Yb metal target into methanol followed by buffer-gas cooling with $\sim$2~K He. The beam is extracted toward a cooling region. A magnetic field is applied at an angle to the laser polarization in order to remix dark states and aid in the Sisyphus cooling. Vibrational repumping in a clean-up region enhances the LIF detection downstream. (b) Spin-rotation (SR) structure typical in the $\tilde{X} \, ^2\Sigma^+$ manifold (not to scale). Each laser indicated in (a) has sidebands at the relevant SR splitting imprinted on it. Note the unusual ordering due to the SR constant, $\gamma$, being negative. (c) Measured values of $\gamma$ for low-lying vibrational states in the $^{174}$YbOH $\tilde{X}\,^2\Sigma$ state. Measurements of the $\tilde{X}(000)$~\cite{Nakhate2019} and $\tilde{X}(100)$~\cite{Steimle2019} spin-rotation constants were previously reported, while measurements for the $\tilde{X}(200)$ and $\tilde{X}(02^00)$ state are reported for the first time here.}
\end{figure}

Figure~\ref{fig:fig-Setup}(a) shows a schematic diagram of the experimental apparatus. YbOH molecules are produced in a cryogenic buffer-gas beam (CBGB)~\cite{hutzler2012buffer,Barry2011}, the essential approach used in all molecular laser cooling experiments.  A cryogenic cell is held at $\sim 2$~K and filled with $^4$He buffer-gas. Hot methanol gas ($\sim 250$~K) is flowed into the cell through a thermally isolated capillary. Laser ablation of Yb metal, followed by a chemical reaction between the Yb atoms and methanol, produces YbOH molecules that are cooled by the He buffer gas. YbOH molecules entrained in the He buffer gas are then extracted into a beam. A typical He flow of 3~sccm is used and the molecules are extracted from the cell through a 16~mm $\times$ 2.4~mm slit (vertical $\times$ horizontal). A typical YbOH CBGB contains $\sim10^9$ YbOH molecules in the $N^{\prime\prime}=1$ rotational level, as measured via absorption spectroscopy. The mean forward velocity is $v_f\sim 90$~m/s and the transverse velocity spread is $v_\perp \sim 15$~m/s. A $2.7$~mm $\times \, 3$~mm aperture placed 20~cm downstream from the cell collimates the molecular beam to an effective transverse temperature of $T_\perp \sim20$~mK.

Just after the aperture, the molecules enter a cooling region of variable length up to $\ell_\text{int} \sim 50$~mm (interaction time $t_\text{int} \sim 500 \, \mu$s). The cooling region contains light at four different wavelengths. The main photon cycling transition is $\tilde{A} \, ^2\Pi (000) \leftarrow \tilde{X} \, ^2\Sigma^+(000)$ (577~nm)~\cite{Melville2001,Steimle2019}. Significant optical pumping into higher lying vibrational states occurs. Linear triatomic molecules possess three vibrational modes: the symmetric stretch, antisymmetric stretch, and doubly degenerate bend. We label the vibrational levels using the notation $(v_1 v_2^\ell v_3)$, where $v_1$ indicates the number of quanta of excitation in the symmetric stretching mode, $v_2$ the quanta in the bending mode, and $v_3$ the quanta in the antisymmetric stretching mode. $\ell$ labels the excitation of nuclear orbital angular momentum in the bending mode. In the present work, dominant decays are to the $\tilde{X} \, ^2\Sigma^+ (100), (200)$, and $(02^0 0)$ levels, and vibrational repumping is required to return this population to the main cooling cycle. Note that vibrational angular momentum selection rules strongly limit decays to the $(01^10)$ bending mode~\cite{HerzbergVol3}. Thus, the cooling region also includes light to drive $\tilde{A} \, ^2\Pi (000) \leftarrow \tilde{X} \, ^2\Sigma^+(100)$ (595~nm, ``first repump"), $\tilde{A} \, ^2\Pi (000) \leftarrow \tilde{X} \, ^2\Sigma^+(02^0 0)$ (599~nm, ``second repump"), and $[17.73] (000) \leftarrow \tilde{X} \, ^2\Sigma^+(200)$ (600~nm, ``third repump")~\footnote{The $[17.73]$ state is an electronically excited state present in YbOH. It behaves mostly like a $^2\Pi_{1/2}$ state and, importantly for the laser cooling scheme, decays predominantly to the $\tilde{X} \, ^2\Sigma^+ (000)$ and $(100)$ levels.}.

The magnetically-assisted Sisyphus effect has been described in detail previously~\cite{Emile1993}. In brief, spatially-varying light shifts, optical pumping into dark magnetic sublevels, and remixing by a static magnetic field lead to an effective friction force which can be much larger than typical Doppler cooling forces. Due to the angular momentum structure of the YbOH transitions used here, we expect cooling at blue detuning and heating at red detuning. To produce the cooling force, the main and first repump laser beams propagate perpendicular to the molecular beam making 12 round-trip passes between mirrors to create a standing wave. Each beam has diameter $\sim$4~mm. A variable magnetic field is present in the cooling region to provide the remixing between dark and bright states necessary for the magnetically-assisted Sisyphus effect~\cite{Emile1993,Berkeland2002}. The field is oriented at 45 degrees to the cooling light's polarization axis to ensure near-optimal remixing. In the case of Doppler cooling (as opposed to the Sisyphus configuration) the standing wave is purposely destroyed by misaligning the retro-reflected beams, but the magnetic field is left on to remix any transient dark states. The second and third repump laser beams, of diameter 6~mm, enter the cooling region through separate fibers, cover the cooling region, and do not form standing waves. For both Doppler and Sisyphus cooling, the light from each laser is split into two frequency components separated by the spin-rotation (SR) splitting of the $N^{\prime \prime}=1$ state; these address the $P_1(N^{\prime \prime}=1)$ and $^P Q_{12}(N^{\prime \prime}=1)$ lines~\cite{HerzbergVol1} (see Fig.~\ref{fig:fig-Setup}(b)). These transitions are rotationally closed, and have been used in all previous molecular laser cooling experiments.

The main cooling light at 577~nm is generated by the second harmonic of a Raman fiber amplifier, while all repumping and imaging wavelengths are generated by cw dye lasers. Up to 150~mW of main cooling light, 100~mW of $(100)$ repumping light, and 50~mW of $(200)$ and $(02^00)$ repumping light are incident on the molecules from each direction. The SR structure in the $\tilde{X}\,^2\Sigma^+$ state, which arises from interaction with $\Omega = 1/2$ excited electronic states, is unusual because the splitting scales quickly with vibrational level (see Fig.~\ref{fig:fig-Setup}(c)). The same effect has been observed in YbF and is attributed to competing contributions from the $\tilde{A} \,^2\Pi_{1/2}$ vibrational states and vibrational levels of $\Omega=1/2$ excited electronic states arising from an Yb$^+(f^{13})$ configuration~\cite{Dolg1992, Sauer1996, Sauer1999, Lim2017YbFSpectroscopy, Nakhate2019}. To address both SR components, we pass each laser beam through an acousto-optic modulator (AOM) which adds a frequency sideband at the appropriate rf interval. Hyperfine splittings in $^{174}$YbOH are negligible due to the large distance between the hydrogen nucleus and the predominantly metal-centered valence electron.

After the cooling region, the molecules travel 40~cm in free flight and then encounter clean-up and imaging regions. In the ``clean-up" region only the (200) and $(02^0 0)$ repumping light is present in order to return any population in excited vibrational states back to the lowest vibrational levels in preparation for imaging via laser-induced fluorescence (LIF). In the imaging region, the molecular beam is imaged with an EMCCD by driving the $\tilde{A}(000) \leftarrow \tilde{X}(000)$ and $\tilde{A}(000) \leftarrow \tilde{X}(100)$ transitions and collecting LIF at 577~nm. The imaging light contains 10~mW of main line and 40~mW of $(100)$ repumping light. The detection light is always resonant and retroreflected in order to avoid velocity-dependent artifacts from appearing in the images. The imaging system is calibrated to ensure that the magnification and collection efficiency are roughly uniform over the entire field of view of 22~mm.

\begin{figure}[t!]
\begin{centering}
\includegraphics[width=.8\columnwidth]{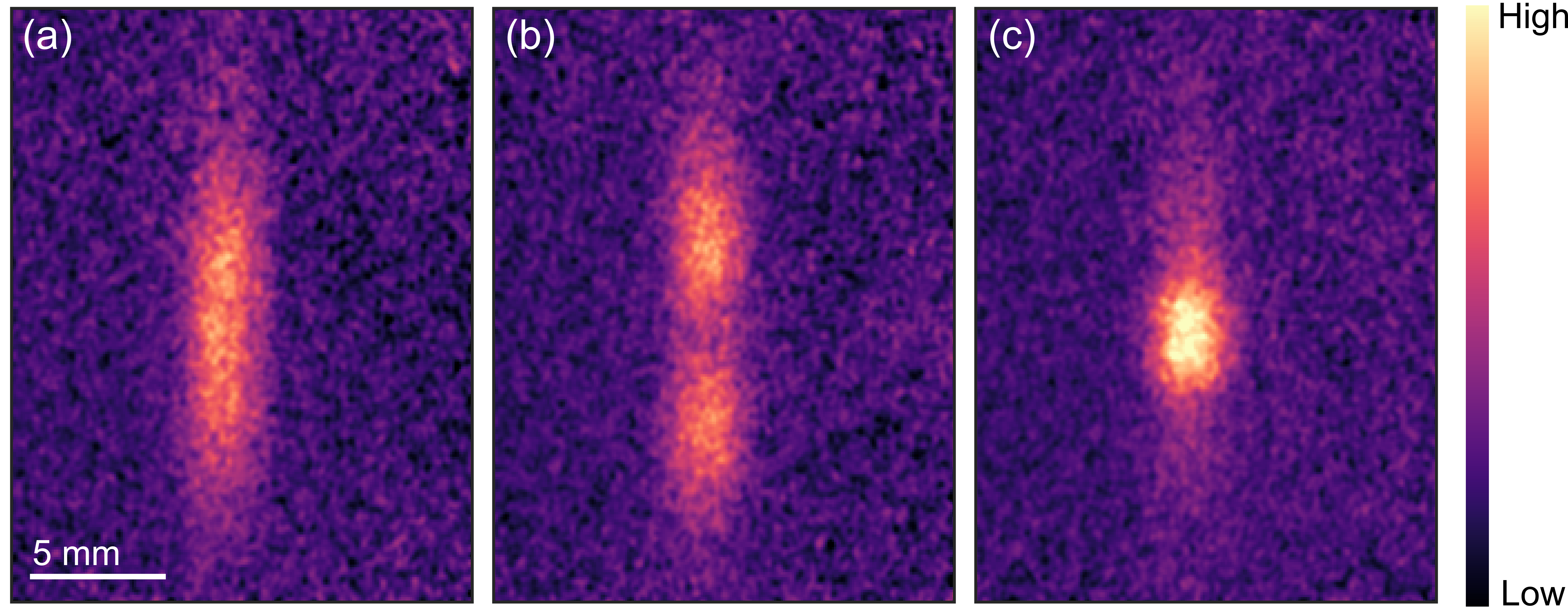}
\par\end{centering}

\caption{\label{fig-2DImages} Images of the YbOH molecular beam under Sisyphus conditions at various detunings. The images represent the average over 20 experimental cycles and are plotted on the same color scale. The molecular beam travels from left to right while the detection laser beams are in the vertical direction. (a) On-resonance light in the cooling region. (b) Red-detuned light at $\Delta = -1.8\Gamma$, heating the molecules away from low velocities. (c) Blue-detuned light at $\Delta=+1.8\Gamma$, leading to substantial cooling and enhancing the on-axis density relative to the on-resonance case.
}
\end{figure}

\section{Results and Discussion}
A typical set of EMCCD images under Sisyphus conditions, expected to give heating at red detuning and cooling at blue detuning~\cite{Emile1993}, is shown in Fig.~\ref{fig-2DImages}. Figure~\ref{fig-2DImages}(a) shows a typical molecular beam image when the lasers in the cooling region are all tuned to resonance. At the highest intensities  ($I/I_{\rm{sat}} \sim 70$) roughly 40\% of the molecules remain for detection. The loss is attributed to decays to both the $\tilde{X}(300)$ and $\tilde{X}(01^10)$ levels, whose precise energies are unknown. The branching ratios to both states are known, however, and the loss allows us to calibrate the number of photons scattered per molecule. Using the measured branching ratios, $1.0(0.25) \times 10^{-3}$ to the $\tilde{X}(300)$ level and $0.8(3) \times 10^{-3}$ to the $\tilde{X}(01^10)$ level~\cite{SteimleInPrep}, we determine that the molecules scatter an average of 500$^{+300}_{-75}$ photons. The scattering rate is measured independently, via images of fluorescence over the molecules' path, to be $\Gamma_\text{sc} \approx 1.5 \times 10^6$~s$^{-1}$. This is consistent with the estimated number of photons scattered as determined by loss to dark states.

\begin{figure}[t]
\begin{centering}
\includegraphics[width=0.9\columnwidth]{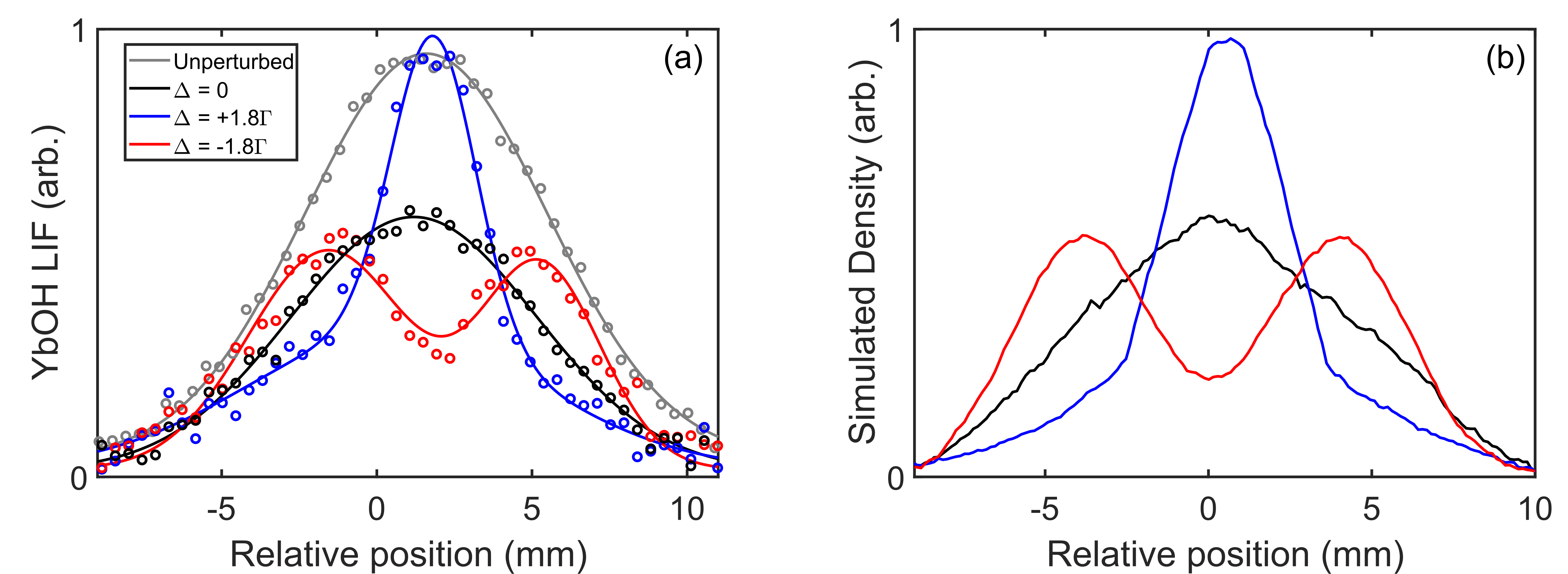}
\par\end{centering}

\caption{\label{fig-IntegratedTraces} (a) Integrated images under Sisyphus cooling conditions. We integrate over the central portion of each EMCCD image and plot the resulting density distribution in the direction of the cooling light. The relatively flat top of the cooled beam is indicative of cooling to the collimating aperture's width. Lines represent Gaussian fits as described in the text. (b) Simulated spatial distribution in the imaging region for the same experimental parameters. The Monte Carlo simulation is described in detail in the text.
}
\end{figure}

In Fig.~\ref{fig-2DImages}(b) (and Fig.~\ref{fig-2DImages}(c)), the molecular beam is imaged after passing through a cooling region with light detuned by $1.8\Gamma$ to the red (and blue) of resonance. ($\Gamma = 1/\tau = 2\pi \times 9$~MHz.) At $\Delta = -1.8\Gamma$ there is a clear heating feature indicated by expulsion of population from the central beam axis. At $\Delta = +1.8\Gamma$ the molecules are cooled and collimated, which results in an increase in on-axis density. Integrating across the central region of these images results in Fig.~\ref{fig-IntegratedTraces}(a). We note that Sisyphus heating concentrates population at specific non-zero velocities. This is because Sisyphus heating is operative at red detuning, but only over a small range of velocities. The Doppler forces become dominant beyond the capture velocity of the Sisyphus force, and population accumulates at the velocity for which these forces balance. This allows us to estimate a value for the capture velocity of the Sisyphus effect of approximately 1~m/s at $I/I_\text{sat} \sim 65$. As discussed below, this is in good agreement with our theoretical model.

In order to interpret the integrated images we fit the transverse beam profiles to a set of Gaussians. The resonance condition molecular beam profile (Fig~\ref{fig-IntegratedTraces}(a)) is first fit to a single Gaussian. In all experimental conditions it is found that the width of the beam is unchanged if light is present in the cooling region but tuned to resonance; only the overall amplitude of the beam signal changes in this case. Next, the cooled beam profile (Fig.~\ref{fig-IntegratedTraces}(b)) is fit to the sum of two Gaussians: one whose width is constrained to match the uncooled signal and another whose width is allowed to vary. In this way we capture the fact that not all molecules in the experiment are below the transverse capture velocity of the Sisyphus force. The relative areas under the ``cooled" and ``uncooled" portions give a measure of the fraction of molecules within the Sisyphus capture range. In our data the cooled portion of the beam typically contains up to 60\% of the molecules. This is consistent with the capture velocity estimated above and the expected velocity distribution produced by the collimating aperture. At maximum intensity, the Sisyphus cooling configuration reduces the FWHM of the cooled molecular beam from 8.5(5)~mm to 3.0(4)~mm.

\begin{figure}[t]
\begin{centering}
\includegraphics[width=\columnwidth]{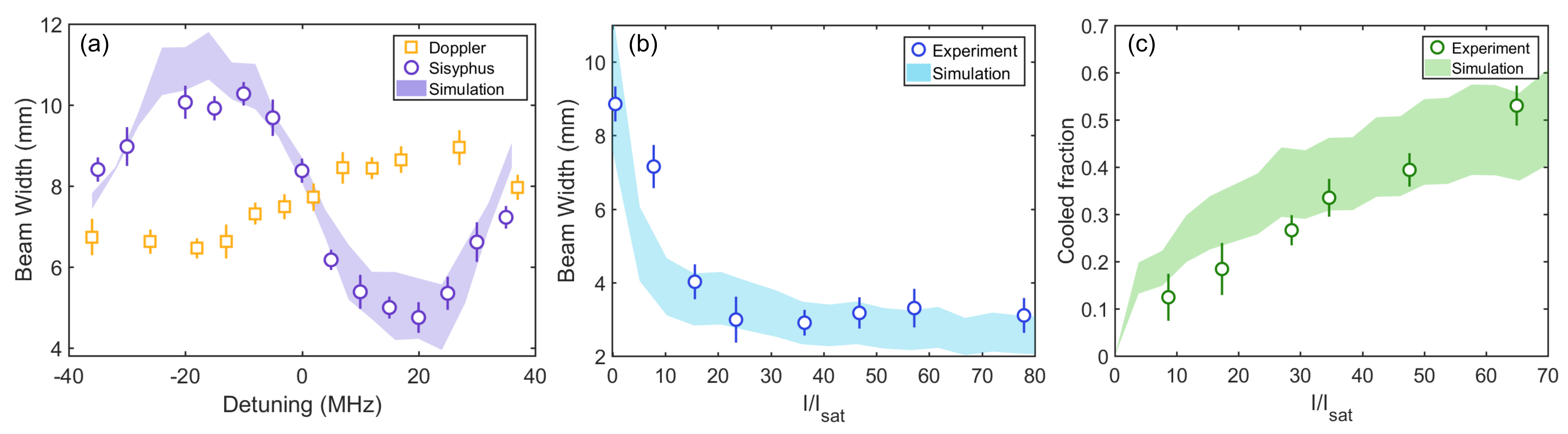}
\par\end{centering}

\caption{\label{fig-DopplerAndIntensity} (a) Comparison of Doppler and Sisyphus cooling as a function of detuning. Data taken at an intensity of $I/I_\text{sat} \sim 15$. The magnetically-assisted Sisyphus effect has the opposite detuning dependence and higher cooling efficiency when compared to Doppler cooling. (b) Dependence of the molecular beam width on Sisyphus cooling light intensity, at detuning $\Delta = 1.5\Gamma$. The beam width initially decreases as the cooling intensity is increased, before saturating to a width set by the collimating aperture. Temperature resolution is limited to $\sim$600~$\mu$K in direct fits of the width. (c) Fraction of the beam within the ``cold" portion of the beam, as determined by our two-Gaussian fit. Data taken at $\Delta = 1.5\Gamma$. In all plots, error bars on experimental points come from the spread in repeated measurements. Shaded regions represent the spread in simulation outputs when the simulated laser beam overlap in the cooling region is varied, as described in the main text.
}
\end{figure}

By misaligning the retro-reflected laser beams to remove the standing wave condition in the cooling region, we are also able to exclusively study Doppler cooling of the molecular beam.  The fitted beam widths as a function of laser detuning are shown in Fig.~\ref{fig-DopplerAndIntensity}(a), which are collected at the same magnetic field ($B \sim 1.5$~G) and with the same number of photons scattered ($\sim 500$) as for Sisyphus cooling. In contrast to Sisyphus cooling, the Doppler scan shows cooling at red detuning and heating at blue detuning. The Doppler cooling and heating are efficient over a large range of frequencies due to the power broadening at the high intensities used in this study. The overall magnitude of the cooling and heating forces are significantly smaller in the Doppler configuration than in the Sisyphus configuration. This is due to the large light shifts present in the molasses, as described below.

In the Sisyphus configuration, we expect the transverse temperature to decrease as the standing wave intensity is increased, as shown in Fig.~\ref{fig-DopplerAndIntensity}(b). The data shows a clear saturation of the minimum beam width at a size set by the collimating aperture, i.e. the molecules expand negligibly during the 40~cm of free flight following this aperture. In order to fit the images to a transverse temperature, we follow Ref.~\cite{kozyryev2016Sisyphus} in comparing the beam width after the cooling region to a Monte Carlo model of ballistic expansion. This yields a minimum temperature of 500$^{+100}_{-450}$~$\mu$K. The temperature has large and asymmetric error bars because the cooled beam is close to the size of the 3~mm collimating aperture, so fitting the cooled cloud size yields very limited temperature resolution. At this temperature, the uncertainty of the fit is comparable to the central value, such that we can only place an upper limit on the temperature. We therefore interpret the extracted temperature as a bound, $T_\perp < 600$~$\mu$K.  Because there is no fundamental temperature limit to the Sisyphus cooling near this scale, the actual temperature is likely much lower; this is discussed in more detail below, where we present the results of calculations using the optical Bloch equations. Figure~\ref{fig-DopplerAndIntensity}(c) shows the fraction of the molecular beam that has been cooled. The value increases with intensity to approximately 60\% at $I/I_\text{sat} \sim 65$, and is not yet saturated at the intensities explored in this work. Given the spread of velocities in the unperturbed molecular beam, this value of the cooled fraction is consistent with the estimated capture velocity of $\sim$1~m/s. From this cooled fraction, we estimate that around $5 \times 10^4$ molecules are in the ultracold part of the distribution. Accounting for the loss to dark states and reduction in transverse temperature, this corresponds to an increase in phase-space density (PSD) in one dimension of a factor of 6~\footnote{Here, the phase-space density is given by $n \lambda_{dB,x} \lambda_{dB,y} \lambda_{dB,z}$ where $n$ is the number density and $\lambda_{dB,i} = h/\sqrt{2 \pi m k_B T_i}$ is the de Broglie wavelength for molecules with mass $m$ and temperature $T$. Because the cooling is only in one dimension, PSD scales with the square root of the temperature rather than as $T^{3/2}$.}.

We have constructed a detailed model of the laser cooling forces based on Refs.~\cite{Emile1993, Devlin2016, Devlin2018}. Because the SR splittings are all large compared to the detunings in the experiment, we model the force on a molecule as the average of the forces on isolated $J^\prime=1/2 \leftrightarrow J^{\prime\prime}=3/2$ and $J^\prime=1/2 \leftrightarrow J^{\prime\prime}=1/2$ subsystems ($\vec{J}=\vec{N}+\vec{S}$), weighted by the number of states. We account for the presence of excited vibrational levels in the $\tilde{X}$ manifold, following Ref.~\cite{norrgard2015sub}, by scaling the saturation intensity to account for the multi-level systems. For each subsystem we solve generalized optical Bloch equations to compute the time evolution of the density matrix as the molecule is dragged at constant velocity through a standing wave. Once the density matrix has reached a periodic steady state, we compute the average force and scattering rate as a function of velocity. We repeat this simulation at a range of magnetic fields, detunings, intensities, and polarizations.

Using the theoretically generated force profiles, we simulate the propagation of molecules through the experimental setup including the full three-dimensional structure of the molecular beam and the transverse profiles of each laser beam, accounting for imperfect laser beam overlap in the cooling region. We incorporate random decays into dark vibrational states and momentum diffusion from spontaneous emission. The intensity, magnetic field, number of photons scattered, polarization, and detuning are all set to the experimentally measured parameters. A representative set of simulated beam profiles is shown in Fig.~\ref{fig-IntegratedTraces}(b) in comparison to the experimental data. In order to assess the uncertainty in the simulation, we leave all parameters fixed to the measured values but vary the assumed laser beam overlap in the cooling region (the most poorly determined parameter), effectively scaling the cooling force up or down by a factor of $\sim$2. Error bars associated with theory curves show the spread of predicted values under this variation (see Fig.~\ref{fig-DopplerAndIntensity}). 

Though the temperature sensitivity of our direct measurements is limited by the collimating aperture width as described above, we extrapolate to lower temperatures using this theoretical simulation. First, we validate the model at higher temperatures, where excellent agreement is seen between the measured and predicted beam widths and cooled fractions (see Fig.~\ref{fig-DopplerAndIntensity}). At saturation parameters $>20$ where the experimental sensitivity has fallen off, our simulation predicts the temperature continues to drop as would be expected intuitively. At the highest intensity realized in the experiment, the model predicts temperatures as low as 10$^{+20}_{-5}$~$\mu$K, which is below the Doppler limit (200~$\mu$K) and approximately 60 times the recoil limit (150~nK) for YbOH. This temperature is comparable to those achieved in the recent work on transverse Sisyphus cooling of YbF molecules~\cite{Lim2018}, where a similar extrapolation was required in order to assign the very low transverse temperatures observed. Our model also allows us to compute a capture velocity for the Sisyphus cooling of $v_c = 0.8(2)$~m/s. This is in good agreement with the value determined from the data above and also consistent with the observed fraction in the ``cooled" part of the distributions. Using the minimum predicted temperature would yield a PSD increase of a factor of 40, as compared to the more conservative estimate based on the temperature limit included above.

\section{Conclusion}
In summary, we demonstrate Sisyphus and Doppler laser cooling of the polyatomic radical $^{174}$YbOH. Under Sisyphus conditions, the transverse temperature of the YbOH beam is reduced from $\sim$20 mK to $< 600 \, \mu$K with $\sim$500 photons scattered per molecule. Sisyphus cooling is found to be significantly more efficient (per photon scattered) than Doppler cooling, as expected due to the large light shifts induced by the near-resonant standing waves~\cite{Emile1993,kozyryev2016Sisyphus}. We compare our results to simulations based on the optical Bloch equations and find agreement over a range of parameters. Validating this model is important as it can be used to guide future experiments to cool and trap YbOH and other molecules sensitive to BSM physics. Due to the expected optical pumping to unaddressed vibrational states, the on-axis density of the cooled molecular beam is approximately equal to the unperturbed beam. The decrease in temperature leads to an increase in the PSD around a factor of 6 under the very conservative assumption of 600~$\mu$K for the transverse temperature of the cooled beam.

Our results are a proof-of-principle demonstration toward further laser cooling of YbOH samples to temperatures low enough for optical trapping. In addition, the transverse cooling demonstrated here could be extended to two dimensions, increasing the number of molecules in an optical trap. The temperature limits determined here are set by the number of photons scattered per molecule, which in turn is set by the number of vibrational repump lasers used in the experiment. This could be improved in future experiments. Measurements of the vibrational branching ratios to $\tilde{X}(300)$ and $\tilde{X}(01^10)$ indicate that adding these two additional repumping lasers will increase the number of photon scatters per molecule to $\sim$3,000 before decay to other unaddressed vibrational levels (with higher vibrational quantum numbers) becomes a limiting factor~\footnote{We note that emission to higher bending modes, where allowed by symmetry (e.g. to the $\tilde{X}(12^00)$ state) may also become relevant at this level.}. Adding two or three more repumping lasers would bring this figure to $>10,000$ photons~\cite{Sharp1964,Nicholls1981}, sufficient to produce a magneto-optical trap of YbOH. Transverse Sisyphus cooling could also contribute directly to improvements in beam-based precision measurements by increasing the number of molecules probed while also allowing longer beam lines, and thus longer coherence times. Straightforward extension of the technique to $^{173}$YbOH would be useful for proposed nuclear magnetic quadrupole moment measurements~\cite{kozyryev2017PolyEDM,Maison2019}.

In closely related work, Ref.~\cite{Jadbabaie2019} recently demonstrated ${\sim}10\times$ enhanced production of YbOH by exciting Yb atoms to the metastable $^3P_1$ state inside a buffer-gas cell, greatly enhancing the flux in a buffer-gas beam. By combining such an enhanced source with previously demonstrated laser slowing, magneto-optical trapping, and transfer to an ODT, we estimate that trapping of $\sim10^5$ molecules in an eEDM experiment is feasible~\footnote{This estimate is based on previously reported efficiencies for slowing and trapping diatomic molecules from CBGBs~\cite{anderegg2017CaFMOT,Truppe2017b,Cheuk2018lambda}.}. This would lead to an eEDM sensitivity surpassing the current limit of $<1.1 \times 10^{-29} \, e\text{ cm}$~\cite{ACME2018}. Additional technical improvements, such as beam focusing~\cite{Kaenders1995}, transverse confinement~\cite{DeMille2013}, and ``few photon" slowing methods~\cite{Lu2014,Fitch2016}, could further increase the sensitivity.

\section{Acknowledgments}
We are grateful to the PolyEDM collaboration, the Hutzler group at Caltech, and Amar Vutha for valuable advice. We thank Louis Baum and Ivan Kozyryev for helpful discussions. This work was supported by the Heising-Simons Foundation. BLA acknowledges support from the NSF GRFP, and ZDL partial support by the Keck Foundation.

\bibliography{YbOHSisyphus_library}

\end{document}